# Picosecond Transient Thermoreflectance for Thermal Conductivity Characterization

Jihoon Jeong[1], Xianghai Meng[1], Ann Kathryn Rockwell[2], Seth R Bank[2], Wen-Pin Hsieh[3], Jung-fu Lin[4,5*] and Yaguo Wang[1,5,*]

[1]*Department of Mechanical Engineering and* [2]*Department of Electrical and Computer Engineering, The University of Texas at Austin, USA*

[3]*Institute of Earth Sciences, Academia Sinica, Nankang, 11529 Taipei, Taiwan*

[4]*Jackson School of Geosciences, and* [5]*Texas Materials Institute, The University of Texas at Austin, USA*

*Author to whom correspondence should be addressed
: yaguo.wang@austin.utexas.edu & afu@jsg.utexas.edu

**Abstract**

We developed a picosecond transient thermoreflectance (ps-TTR) system for thermal property characterization, using a low-repetition rate picosecond pulsed laser (1064 nm) as heating source and a 532 nm CW laser as probe. Low-repetition rate pump eliminates the complication from thermal accumulation effect. Without the need of a mechanical delay stage, this ps-TTR system can measure thermal decay curve from 500 ps up to 5 μs. Three groups of samples are tested with this ps-TTR system: bulk crystals (Si, GaAs and sapphire); $MoS_2$ thin films (157 nm ~ 900 nm); InGaAs random alloy and GaAs/InAs digital alloy (short period superlattices). Analysis of the thermoreflectance signals show that this ps-TTR system is able to measure both thermal conductivity and interface conductance. The measured thermal conductivity values in bulk crystals, $MoS_2$ thin films and InGaAs random alloy are all consistent with literature values. Cross-plane thermal conductivity in $MoS_2$ thin films do not show obvious thickness dependence, suggesting short phonon mean free path along cross-plane direction. Thermal conductivities of GaAs/InAs digital alloys are smaller than InGaAs random alloy, due to the efficient scattering at interfaces. We also discuss the advantages and disadvantages of this newly developed ps-TTR system comparing with the popular time-domain thermoreflectance (TDTR) system.

Keywords: Thermoreflectance, cross-plane thermal conductivity, interfacial thermal conductance, picosecond

**INTRODUCTION**

As developing high-power-density micro-/nano-scale size electronics, thermal management becomes critical. One fundamental problem of thermal management is characterization of thermophysical properties. In the past decade, noncontact thermoreflectance techniques have been invented in various forms, such as time-domain thermoreflectance (TDTR) [1], [2], frequency-domain thermoreflectance (FDTR) [3]-[5], transient thermoreflectance (TTR) [6], transient thermal grating (TG) [7], [8] and transient grating imaging [9]. Even though different in experimental configurations and analytical models, the common feature of these techniques is the use of a strong pump laser as the heating source to elevate the surface temperature and a weak probe laser to monitor the surface temperature change. TDTR employs a high-repetition-rate laser (e.g. 80 MHz) in conjunction with chopping pump beam at high frequency with Electro-optic Modulator (EOM). The time delay of TDTR is controlled with a mechanical delay stage and signal is acquired with a lock-in amplifier. Experimental signals of TDTR have good sensitivity to thermal conductivities of nanoscale thin films and thermal interface resistance/conductance [2]. Because the duration between pulses is not enough for the system to recover its original thermal state [10], the effect of heat accumulation needs to be considered for data analysis. Use of the mechanical delay stage limits the total detection time to several nanoseconds at a high cost. FDTR technique uses either pulse lasers or continuous wave (CW) lasers for both pump and probe and measures signal in frequency domain, instead of time domain, over a wide range of modulation frequency. FDTR can measure the thermal conductivity and heat capacity of a sample simultaneously if the thermal diffusivity is larger than $3\times10^{-6}$ $m^2$/s [3]. However, it also means that it is challenging to investigate low-κ materials using FDTR. Both TDTR and FDTR can also measure in-plane thermal conductivity by carefully varying spot sizes and modulation frequencies [3], [11].

We developed a picosecond transient thermorefletance system (ps-TTR) that utilizes a low-repetition rate picosecond pulsed laser as pump and a CW laser as probe. This ps-TTR has several advantages that makes it appealing for thermal property characterization. (1) It detects a thermal decay signal excited by a single pulse, and the time between pulses is 5 μs and long enough for the sample surface temperature to return to its original state. No heat accumulation effect is expected. (2) Using a CW probe eliminates the need of a mechanical delay stage. The delay time could be as long as 5 μs, only limited by the laser repetition rate. (3) Fast data acquisition could be achieved with a digital oscilloscope [6]. (4) Unlike the nanosecond laser TTR that is mainly sensitive to thermal conductivity in bulk materials, this ps-TTR system can measure thermal conductivity in nanostructures as well as interfacial thermal conductance. A grating imaging technique, recently developed in our group [9] and similar to the heterodyned transient grating technique [7], [8], can be incorporated into this ps-TTR system for in-plane thermal conductivity measurement. With this newly developed ps-TTR system, we firstly examine thermal conductivities in three bulk materials, including Si, GaAs, and sapphire; then tested thin-film samples including $MoS_2$ thin-films over a wide range of thicknesses, and digital alloy samples of GaAs/InAs. Finally, we discuss the advantages/disadvantages of this ps-TTR method comparing with the popular TDTR technique.

**EXPERIMENTAL**

The schematic layout of this ps-TTR system is shown in Fig. 1. A picosecond pulsed laser (Coherent Talisker Ultra 532-8, 1064 nm central wavelength, 15 ps pulse width, and 200 kHz repetition rate) is used as the pump and a Nd:YAG laser (532 nm, CW, Coherent Verdi V6) as the probe. A 10x objective lens is used to focus both the pump and probe beams onto the sample surface, and the reflected probe beam is collected by a Hamamatsu C5658 silicon avalanche photodiode with 1 GHz bandwidth (pulse response full width at half maximum, FWHM: 500 ps). The signal traces are recorded on an oscilloscope with 4 GHz bandwidth (TDS 7404, Tektronics). The pump spot diameter ($1/e^2$) is about 125 μm while the probe diameter is about 10 μm. For all the samples measured, the upper limit of thermal penetration depth is less than 11 μm ($d_p = \sqrt{\alpha t_p}$; estimated using the thermal diffusivity of Silicon, $\alpha=9.34\times10^{-5}$ $m^2/s$ and the longest fitting time range $t_p=300$ $ns$), which is much smaller than the pump spot diameter. These two conditions ensure that cross-plane thermal transport is dominant and a simple 1-D thermal conduction model can be used to extract the cross-plane thermal conductivity. Metal thin films are deposited onto all sample surfaces as heat transducers to enhance thermoreflectance signal and to maintain a linear relation between reflectance and temperature change. When choosing the metal transducer, preferred are high thermoreflectance coefficient ($dR/dT$) and low absorbance (1-R) at probe wavelength. High $dR/dT$ value will provide high measurement sensitivity and low absorbance will avoid unnecessary heating from probe laser. Based on these conditions, Au thin film is selected as metal transducer due to its high thermoreflectance value ($\sim 2\times10^{-4}$ $K^{-1}$) and low absorbance ($< 0.3$) at probe wavelength (532 nm) [12].

To simulate the experimental results and extract thermal properties, the 1D thermal diffusion equation in a multi-layer model, including the metal transducer, sample and substrate (when applicable), is solved in time domain with Finite Difference Method [6]:

$$\rho_m c_m \left(\frac{\partial T_m}{\partial t}\right) = \frac{\partial}{\partial z}\left(\rho_m \kappa_m \frac{\partial T_m}{\partial z}\right) + S(z,t) \quad (1)$$

$$\rho_s c_s \left(\frac{\partial T_s}{\partial t}\right) = \frac{\partial}{\partial z}\left(\kappa_s \frac{\partial T_s}{\partial s}\right) \quad (2)$$

where $\rho$ is the density, $c$ is the heat capacity, and $\kappa$ is the thermal conductivity and S is the source term due to pump laser heating. The subscripts $m$ and $s$ denote the metal transducer and sample, respectively. The source term is only considered in the metal layer. The repetition-rate of our picosecond laser system is set at 200 kHz, which is low enough to guarantee that temperature in the material completely relaxes back to its initial value between pulses. A typical thermal relaxation time in our samples is only several hundred nanoseconds, as shown in Fig. 2. (a) and Fig. 3. (c). Therefore, no thermal accumulation effect is considered here. The source term is described as [13]:

$$S(z,t) = \frac{0.94(1-R_{pump})F}{t_p \delta [1-\exp(-\frac{L}{\delta})]} \exp\left[-2.77\frac{(t-2t_p)^2}{t_p^2} - \frac{z}{\delta}\right] \tag{3}$$

where $R_{pump}$ is the sample reflectivity at the pump laser wavelength, $F$ the laser fluence, $t_p$ the pulse width, $\delta$ the optical absorption depth, and $L$ the thickness of the metal layer. The assumption is that all the pump laser energy is absorbed in the metal layer. We used 50 nm or 80 nm gold as transducer. The absorption depth of Au at 1064 nm is only 12 nm, which validates our assumption. The Boundary conditions are given by:

$$-\kappa_m \frac{\partial T_m}{\partial z}\bigg|_{z=L} = -\kappa_s \frac{\partial T_s}{\partial z}\bigg|_{z=L} = G(T_m - T_s)|_{z=L} \tag{4}$$

where $G$ is the interfacial thermal conductance between metal and substrate. If a substrate is used, a third layer will be added and Eq. (2) & (4) will be applied to the substrate and sample/substrate interface, respectively.

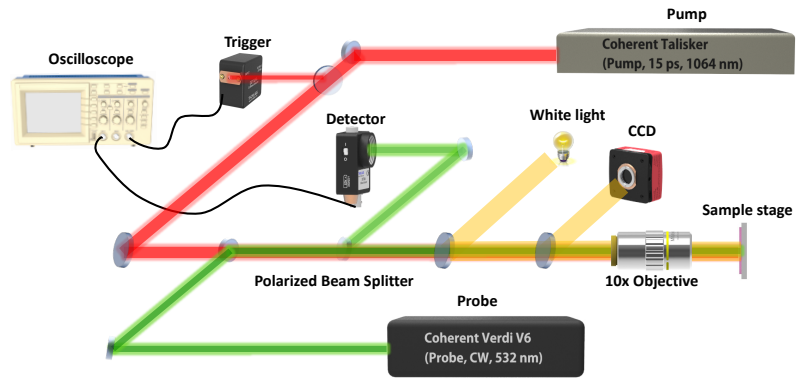

Figure 1. Schematics of ps-TTR set-up.

**RESULTS AND DISCUSSION**

In order to validate this ps-TTR system, three bulk materials are measured first, including sapphire, GaAs and Si. A 50 nm Au thin film is deposited on sample surface by e-beam evaporator. The Au thin film thickness is determined with profilometer (Dektak 6M Stylus Profilometer). Experimental data and the fitting for sapphire is plotted in Fig. 2a. Several parameters need to be considered for the fitting process, including heat capacity ($c$), thermal conductivity ($\kappa$) and thickness ($d$) for both Au layer and the sample, as well as the interface conductance ($G$) between Au and sample. The only unknown parameters are thermal conductivity of the sample and the interface conductance between Au/sample. Reference values are used for all other parameters (listed in Table 1) [14]. Since the delay time of ps-TTR system could be up to 5 µs, it is important to choose a proper time range for fitting, where the temperature decay curve is most sensitive to the parameters sought. As shown in Fig. 2b for sapphire, sensitivity analysis reveals the most sensitive time range for each parameter (See supplemental material for details). The conductance at Au/Sapphire interface, $G$, has the highest sensitivity before 10 ns, whereas thermal conductivity of Sapphire is most sensitive around 100 ns. Here, we employ a multi-parameter fitting process to extract $\kappa_s$ and $G$. The time range chosen for fitting can affect the extracted thermal conductivity values. To minimize the uncertainty caused by this step, we also conducted the fitting with different time ranges. We found fitted values of $\kappa_s$ and $G$ converge when the time range is long enough. (See supplemental material for the case of sapphire). Uncertainty analysis for two parameters including $\kappa_s$ and $G$ is performed using Jacobian matrix between known and unknown parameters [15], [16] [See supplemental material] . Uncertainty values are calculated with 10% uncertainty for $d_{metal}$, and 3% for other parameter, since it is difficult to measure metal film thickness very accurately. As shown in Fig. 2c, the resulted uncertainty for both $\kappa$ ang $G$, reaches their lowest value around 100 ns, and becomes almost constant thereafter. The extracted thermal conductivities are plotted in Fig. 2d along with values reported in literature. The $\kappa_s$ value is 37.4±4.1 W/m·K for sapphire, 60.3±6.8 W/m·K for GaAs and 148.5±15.5 W/m·K for Si. The interface conductance $G$ obtained is 33.3±2.7 MW/m²·K for sapphire, 29.5±2.0 MW/m²·K for GaAs, and 29.4±2.1 MW/m²·K for Si. The error bars in Fig. 2d contain both experimental and fitting uncertainties. Both the measured thermal conductivity [16]-[18] and interface conductance [6], [19]-[22] are consistent with literature values, which validates our ps-TTR system.

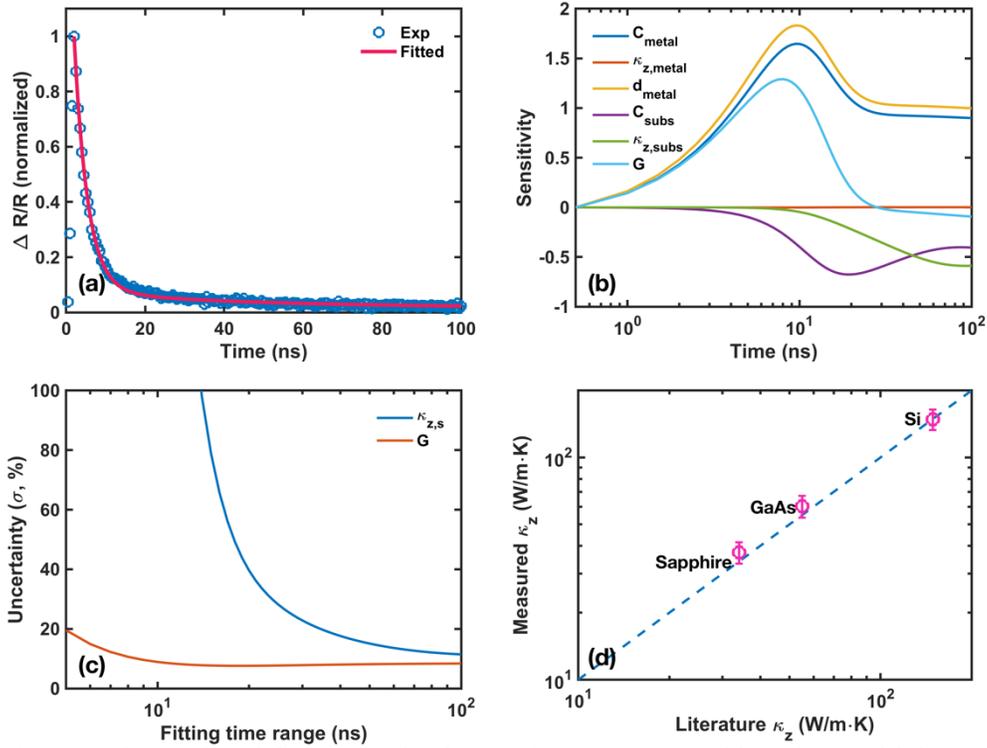

Figure 2.(a) Experimental and fitting results for Au (50 nm) on sapphire, (b) sensitivity results for all parameters related to the fitting process, (c) uncertainties of Sapphire thermal conductivity and interface resistance based on sensitivity results, (d) thermal conductivity of sapphire, GaAs and Si measured with ps-TTR system, along with literature values [16]-[18].

Table 1. Thermophysical properties used in simulation model.

|  | Density (kg/m$^3$) | Heat Capacity (J/kg·K) | Thermal Conductivity (W/m·K) |
|---|---|---|---|
| **Au [14], [23]** | 19300 | 129.1 | 310 |
| **Si [17], [18]** | 2329 | 710 | 149 |
| **GaAs [17], [18]** | 5317 | 350 | 55 |
| **Sapphire [16]** | 3835 | 879 | 34 |
| **MoS$_2$ [16], [24], [25]** | 5060 | 373.5 | 2~5 |

We also utilize this ps-TTR system to measure thermal conductivity in MoS$_2$ thin films, a most studied member in the transition metal dichalcogenides (TMDs) family. MoS$_2$ thin-films are prepared via mechanical exfoliation from bulk crystal (purchased from 2D semiconductors), and then transferred onto SiO$_2$ (90 nm) / Si substrates using non-residual semiconductor tape from UltraTape. Seven samples with different thickness are prepared, including 157, 288, 330, 435, 464, 620, and 900 nm, measured with Atomic Force Microscope (AFM; Asylum MFP-3D). 80 nm Au thin film is used as thermal transducer. For the fitting process, known parameters are the heat capacity ($c_{metal}$), thermal conductivity ($\kappa_{metal}$) and thickness ($d_{metal}$) of Au layer, heat capacity ($c_{MoS2}$) and thickness ($d_{MoS2}$) of MoS$_2$, and heat capacity ($c_{subs}$) and thermal conductivity ($\kappa_{subs}$) of Si substrate, as indicated in Table 1. The three unknown fitting parameters are thermal conductivity of MoS$_2$ ($\kappa_{MoS2}$), interfacial thermal conductance ($G_1$) between Au and MoS$_2$ and interfacial thermal conductance ($G_2$) between MoS$_2$ and Si substrate. $G_2$ includes the 90 nm SiO$_2$ layer on Si substrate. Several steps are taken to obtain accurate fitting values: 1) perform sensitivity test for all parameters to find the most sensitive time range for each unknown parameter; 2) calculate uncertainty for $\kappa_{MoS2}$, $G_1$ and $G_2$ based on the sensitivity results; and 3) select a fitting time range where $\kappa_{MoS_2}$ is most sensitive with minimized uncertainty. Fig. 3 demonstrates the case of a 464 nm MoS$_2$ thin film. Fig. 3a shows an example sensitivity results and Fig. 3b gives uncertainties for $\kappa_{MoS2}$, $G_1$ and $G_2$. Uncertainty values are calculated with 10% uncertainty for $d_{metal}$, and 3% for other parameter. As shown in Fig. 3a, sensitivity for $G_1$ has a peak before 20 ns, and while $\kappa_{MoS2}$ is most sensitive at around 300 ns time range. This is also the time when uncertainty for $\kappa_{MoS2}$ is minimized, as shown in Fig 3b. Therefore, the fitting time range is selected to be 300 ns since our most interested parameter is $\kappa_{MoS2}$. One example of experimental data and the fitting is presented in Fig. 3c. Fig. 3d ~3f show the fitted values of $\kappa_{MoS2}$, $G_1$ and $G_2$. For the 620 and 900 nm samples, $G_2$ values are not displayed, because in thick samples the experimental data is not sensitive to $G_2$. All results do not show obvious trend with MoS$_2$ thickness, and the averaged values are $\kappa_{MoS2}$=4.3±0.7 W/m·K, $G_1$=23.7±3.5 MW/m$^2$·K, and $G_2$=16.0±1 MW/m$^2$·K. The error bars includes both experimental and fitting uncertainty. The average thermal conductivity value is consistent with the reported values in the range of 2 to 5 W/m·K [16], [24]-[26]. So far there is no experimental investigation of thickness dependent cross-plane thermal conductivity in MoS$_2$. Thickness dependence on in-plane thermal conductivity is reported in MoS$_2$ less than 10 layer thick [27]. First-principle calculations in Van der Waals' solids suggest that strong thickness dependent cross-plane thermal conductivity is only expected when phonon mean free path (MFP) is comparable with sample thickness, where boundary scattering dominates over phonon-phonon scattering [26]. There have been many studies for MFP calculation in $MoS_2$ in basal direction, with in-plane phonon MFP reported in monolayer in a range of 10-40 nm [28]-[31]. Usually the cross-plane phonon MFP is much shorter than the in-plane one, and hence much shorter than the thickness of all our $MoS_2$ thin films. This explains why no thickness dependence is observed. Using fs-TDTR technique, modulation frequency dependent cross-plane thermal conductivity has been reported in bulk $MoS_2$, which was explained by different thermal penetration depth varying with modulation frequency [16]. However, according to a recent publication, the modulation frequency dependent thermal conductivity observed in TMDs could be due to nonequilibrium thermal resistance between high frequency optical phonons and acoustic phonons in TMDs [16]. In our measurement, the separation time between pump pulses is 5 μs. From Fig. 3c, the sample has already returned to its ambient state after 300 ns, which means all the phonons have reached an equilibrium state at ambient temperature. Therefore, the measured thermal conductivities using this ps-TTR technique are simply intrinsic values.

Mao et al. predicted the interfacial thermal resistance between metal and $MoS_2$ with first-principles calculations based on the Landauer formalism [32]. The calculated value for Au/$MoS_2$ interface resistance is 58 $m^2$·K/GW, which is equal to 17.2 MW/$m^2$·K for the conductance value. Jiang et al. reported modulation frequency dependent $G_1$ value between Al and $MoS_2$, and the $G_1$ value at the lowest modulation frequency of 0.7 MHz is around 35 MW/$m^2$·K [16]. For Au on graphite, Schmidt et al. reported an interface conductance value of ~35 MW/$m^2$·K [33]. Our measured $G_1$ values are consistent with these literature values.

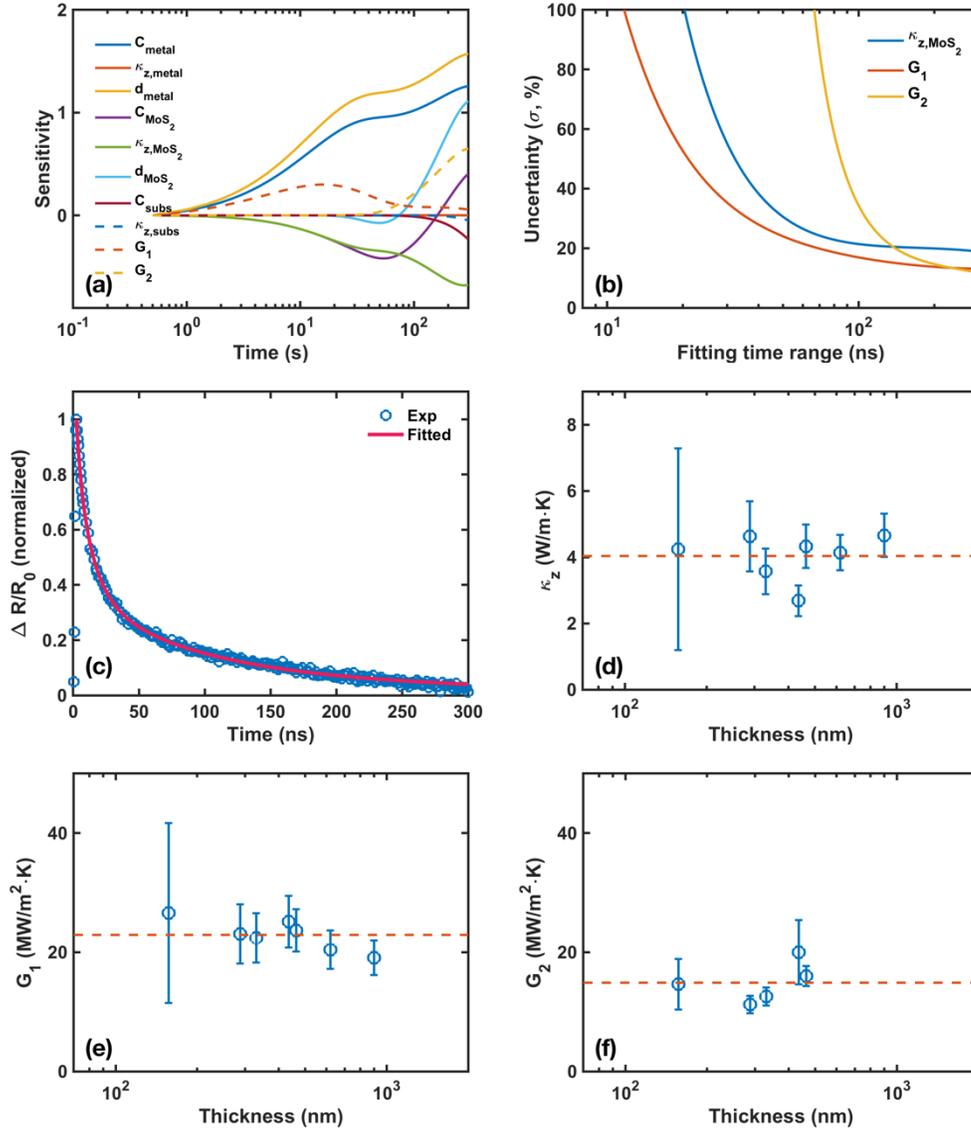

Figure 3. For the case of Au (80 nm) on MoS$_2$ (464 nm) on SiO$_2$ (90 nm) / Si substrate: (a) Sensitivity analysis for all parameters related to the fitting process. (b) Uncertainties of MoS$_2$ thermal conductivity and interface conductance ($G_1$ and $G_2$), based on sensitivity results. (c) Experimental and fitting results. Fitted values of (d) $\kappa_{MoS2}$, (e) $G_1$, and (f) $G_2$ in MoS$_2$ thin-films. For each thickness, five different spots are measured and averaged to calculate the error bars.

The third group of samples tested are alloys, with information listed in Table 2. The actual sample structure is demonstrated in Fig. 4a. All the samples are grown by solid-source molecular beam epitaxy (MBE), lattice matched on semi-insulating InP substrates [34]. Alloy #1 is a conventional random alloy, In$_{0.53}$Ga$_{0.47}$As, where the Ga, In, and As shutters are all open simultaneously, and the composition is controlled by the source temperature. Alloy #2 & #3 are digital alloys, consisting of periodic layers of GaAs and InAs, where the composition is controlled by the duty of cycle of the shutter as opposed to source temperature. The period thickness of these digital alloy is 2.35 nm (3.76 monolayer (ML) GaAs / 4.24 ML InAs per period), and 107 periods in total. While some digital alloys have been shown to possess properties beneficial for photodetectors [35], the material properties of these alloys remain largely unknown. For alloy #2 & #3, an InGaAs buffer layer is firstly grown on the substrate to smooth and prime the surface for high quality growth of the active layers. Alloy #2 also has a cap layer to prevent oxidation of the sample surface after growth.

Table 2. Sample information of InGaAs digital and random alloys.

|  | Alloy #1 | Alloy #2 | Alloy #3 |
| --- | --- | --- | --- |
| Type | Random alloy ($In_{0.53}Ga_{0.47}As$) | Digital alloy (GaAs/InAs) | Digital alloy (GaAs/InAs) |
| Alloy thickness | 250 nm | 250 nm | 250 nm |
| Period thickness | N/A | 2.35 nm | 2.35 nm |
| Cap layer | N/A | 10 nm InGaAs | N/A |
| Buffer layer | N/A | 100 nm InGaAs | 200 nm InGaAs |

* One ML GaAs has a thickness of 0.2827 nm, one ML InAs has a thickness of 0.3029 nm, and one ML InGaAs has a thickness of 0.2934 nm.

For thermal conductivity measurement, we deposited a 50nm Au thin film on all samples. For numerical simulation, we only consider three layers, Au film, alloy and substrate, as shown in Fig.4a. Effects from cap layer and buffer layer are lumped into two interface conductance ($G_1$ and $G_2$). For alloy layer, the values averaged over GaAs and InAs by their weight ratio (47:53) are used for heat capacity (297 J/kg·K) and density (5504 kg/m$^3$), given that for GaAs $\rho$ = 5317.6 kg/m$^3$, c = 350 J/kg·K and for InAs $\rho$ = 5670 kg/m$^3$, c = 250 J/kg·K.

Obtained thermal conductivities from fitting are 3.2 ± 0.6, 2.5 ± 0.3, and 2.5 ± 0.4 W/m·K for Alloy #1, #2, and #3, respectively, as plotted in Fig. 3b. Since reported thermal conductivity value of bulk $In_{0.5}Ga_{0.5}As$ (random alloy) is about 5 W/m·K [36], [37], our measured value for the random alloy is relatively low (3.2 ± 0.6 W/m·K). Considering the thickness of 250 nm, the lower value measured here can be explained by the thickness dependent thermal conductivity when film thickness is shorter than phonon MFP. Koh et. al.

has reported modulation frequency (*f*) dependent thermal conductivity in InGaAs alloy [38], where the modulation frequency determines the thermal penetration depth by a relation of $d_p = \sqrt{\kappa/\pi f \rho c}$. When the thermal penetration depths is shorter than the phonon MFP, ballistic thermal transport starts to play a role, and the measured thermal conductivity is lower than the intrinsic value of bulk. According to reported accumulated thermal conductivities, phonon MFP of GaAs and InAs is on the order of hundreds of nm [39], [40], which is comparable with the thickness of the random alloy. So ballistic heat transport could play a role in our measured thermal conductivity. With thermal penetration depth about 250 nm, the measured thermal conductivity value deduced from [38] is in the range of 3~4 W/m·K, which is consistent with our results (3.2 ± 0.6 W/m·K). Digital alloys (#2 & #3) have lower thermal conductivities, which is a result of very effective phonon scattering at interfaces. Similar phenomena have been observed in many previous publications [41]-[43]. Thermal conductance across two interfaces ($G_1$ for Au/Alloy interface and $G_2$ for Alloy/substrate interface) are plotted in Fig. 4c. Alloy #3 has the smallest $G_2$ value, which is reasonable because it has the thickest buffer layer.

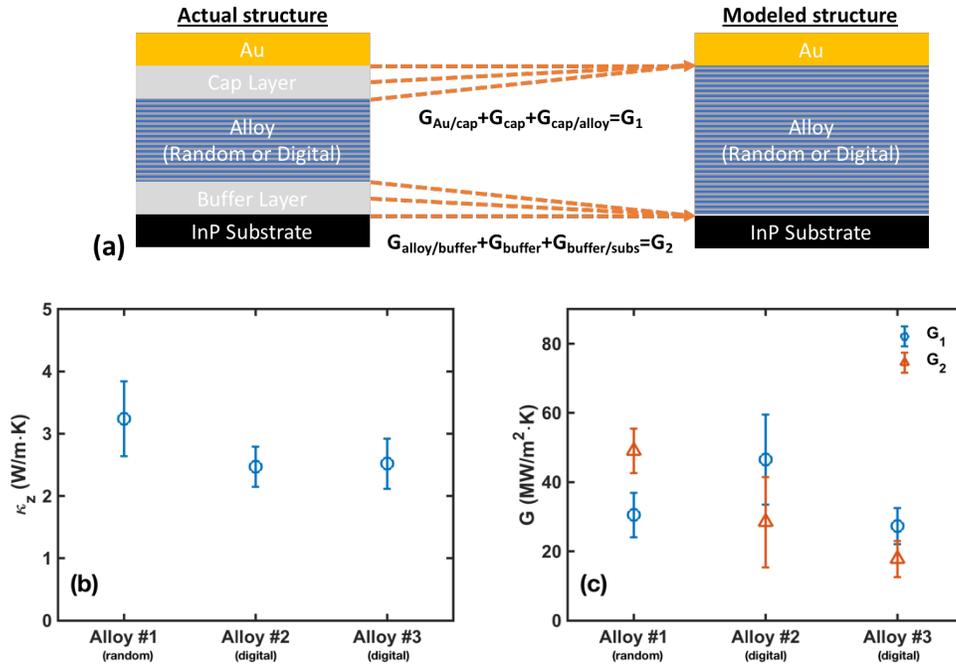

Figure 4. (a) Schematic diagram of actual sample structure (left) and modeled structure for fitting (right). Results of three different alloy samples: (a) thermal conductivity, (b) the first and second interface thermal conductance ($G_1$ & $G_2$).

By characterizing three groups of samples, we have shown that this newly developed ps-TTR system can measure both thermal conductivity and interface resistance with good accuracy. Lastly, we want to comment on the differences between this ps-TTR system and the popular TDTR system. As regard to the experimental set-up, ps-TTR uses a low-repetition rate pump laser, a CW probe laser and a fast photodiode working with an oscilloscope for data acquisition. For TDTR, both pump and probe are pulses form a high-repetition rate laser. Time delay between pump and probe is controlled with mechanical delay stage. Electro-optical modulator (EOM) working with a lock-in amplifier is used for data acquisition. EOM modulation allows control of thermal penetration depth by tuning the modulation frequency. The tunable modulation frequency of TDTR provides an extra free parameter and is useful for many purposes, for example, the phonon MFP spectroscopy. The experimental set up of ps-TTR is much simpler and cost effective. For example, a 1064 nm laser with ~400 ps pulse width can cost as low as $12K (RMPC, Wedge-XF-1064), which is even cheaper than a long range mechanical delay stage. Both systems can be modified to measure in-plane thermal conductivity as well. For TDTR, techniques such as offset laser beams [44] and varying pump or probe laser spot size [11] have been implemented. For ps-TDTR, our recently developed grating imaging technique [9], [45] can be integrated with ease. As regard to the thermal model to extract thermal properties from experimental data, for ps-TDTR we use finite difference method to solve the simple thermal diffusion model numerically. For FDTR, the thermal diffusion model considers heat accumulation effect and is usually solved in frequency domain analytically. Even though numerical simulation is more expensive computationally, it provides a great amount of flexibly to consider effects from laser penetration depth at different wavelength, pulse duration, pulse shape, etc. TDTR was reported to be sensitive to thermal effusivity ($e = \sqrt{\kappa \rho c_p}$) at high modulation frequency and to thermal diffusivity ($\alpha = \kappa/\rho c_p$) at low modulation frequency [46], [47]. Our ps-TTR system is mainly sensitive to thermal effusivity. The detector in our current ps-TTR system has a time resolution of 500 ps, which could be a limiting factor for measuring thermal conductivity in films much thinner than 100 nm.

# CONCLUSION

We developed a picosecond transient thermoreflectance technique for thermal property characterization. This system provides a time delay of several microseconds, only limited by the laser repetition rate, as well as a time resolution of 500 ps, limited by the photodetector response time. The measured thermal conductivity values in bulk crystals, $MoS_2$ thin films and InGaAs random alloy are all consistent with literature values. Our analysis has shown that experimental data taken with this ps-TTR system is sensitive to both thermal conductivity and interface conductance in nanostructures.

# ACKNOWLEDGEMENTS

The authors would like to acknowledge supports from National Science Foundation (NASCENT, Grant No. EEC-1160494; CAREER, Grant No. CBET-1351881; and CBET-1707080). Ann Kathryn Rockwell and Seth R Bank also acknowledge support from NSF Grant No. DMR 1508603, and ARO and DARPA under Contract No. W911NF-17-1-0065.